# Modification of the magnetic tunnel junction properties

A. Filatov,[1] A. Pogorelov,[1] Ye. Pogoryelov[2]

[1] *G.V. Kurdyumov Institute for Metal Physics, NAS of Ukraine, 03142 Kiev, Ukraine*

[2] *University of Gothenburg, Department of Physics, 412 96 Gothenburg, Sweden*

We study the electro-physical properties of the Fe/MgO/Fe magnetic tunnel junctions (MTJ). Sample structures are fabricated on top of glass-ceramic substrates by e-beam evaporation in a relatively low vacuum (~$10^{-4}$ Torr). The influence of the first magnetic layer fabrication conditions on the degradation of the MTJ is explained by the interlayer diffusion. Various models of electro-physical processes in MTJ on polycrystalline substrates are discussed. The current-voltage (I-V) characteristics of the fabricated structures are found to exhibit the region with the negative differential resistance, similar to the one in tunneling diodes. We explain this phenomenon by the formation of excitons in the MgO layer modified by impurities. Obtained results will be useful in the development of MRAM devices containing MTJs and tunneling diodes.

**I. INTRODUCTION**

Traditional methods of fabrication of MTJ structures are based on the state-of-the-art ultra-high vacuum deposition systems with the base pressure of ~$10^{-8}$ Torr.[1,2] Such techniques as molecular beam epitaxy (MBE), electron beam evaporation, magnetron sputtering are typically utilized to grow epitaxial multilayer film structures, including trilayer Fe/MgO/Fe, on single crystal substrates. The apparent advantage of such structures is their crystallographic orientation defined by the substrate, which allows achieving high values of the spin polarization coefficients (up to 70% and above) and the tunneling magneto-resistance (TMR) values better than 150% at room temperature.[3] These parameters are in a good agreement with the existing theoretical results in this field,[4] however require both highly pure materials and high precision technologies with the parameters close to theoretical.

At the same time the mass production of various microelectronic devices and microchips widely involves materials, the composition and structure of which differs from the ideal. For example, polycrystalline corundum (polycor) and glass-ceramics are used as substrates for the microchips.[5] These materials have polycrystalline structure with, correspondingly, large (up to 10 μm for polycors) and small (up to 0.5 μm for glass-ceramic) crystallites. Their surface roughness is better than 30 nm.[6] The morphology of a properly treated surface of such substrates is suitable for fabrication of the multilayer film structures, e.g., Fe/MgO/Fe. These films will have fine-grained structure with the appropriate orientation of the magnetic moment within the film grains.[7] They will not have a well-defined magnetic anisotropy as compared to the films grown on single crystal substrates. Depending on the correlation between the characteristic dimensions of the crystallites and the size of the MTJ such situation may be fair within the entire MTJ. It is expected that in these structures the electro-physical properties will differ from the ones typical for the epitaxial film based MTJs on single crystal substrates. For this the size of the fabricated junction should be much larger than the characteristic domain size. For the fine-grained structures it may be assumed that the domain size is on the order of the grain.

When using traditional techniques for the fabrication of MTJ-structures the existence of impurities degrade their electro-physical properties. At the same time impurities are known to be used to create non-monotonous I-V characteristics in semiconductor microelectronic devices (I-V characteristics with negative differential resistance in tunnel diode). The interest in such characteristics is associated with the possibility of using it to create amplifier or oscillator type semiconductor devices. Theoretically the possibility to obtain the negative differential resistance in MTJ-structure was shown in Ref. 8. In Ref. 9 the authors have studied one of the possible mechanisms of the effect of impurities on the spin-dependent tunneling in ferromagnet/oxide/ferromagnet structures. It was proposed that the existence of impurities may lead to the change of both the value and the sign of TMR. As a result, by introducing impurities into the MTJ-structure one should expect the appearance of new properties of such structure, which can be utilized in development of the new type of spintronic devices.

In this work we study the room temperature electric and magnetic properties of the MTJ-structures fabricated on polycrystalline substrates. Existence of the known impurities in the deposition chamber allows to study their influence on the properties of fabricated film structures.

## II. EXPERIMENT AND RESULTS

The MTJ-structure was deposited by e-beam evaporation of the corresponding materials on the glass-ceramic substrate in the vacuum chamber with the pressure of $10^{-4}$ Torr. Before film deposition the 20x20 mm glass-ceramic substrate was subjected to chemical cleaning. Each layer of the film structure was deposited through a special mask placed on the substrate surface. First Fe layer of 20-30 nm thickness was deposited through the mask having three parallel 150 μm wide slits. Separation between the slits was 1 mm. During the deposition the substrate temperature was kept at 150 °C. Next, the MgO barrier layer was deposited through the mask having a 15x15 mm square cut-out. Before deposition of the third, Fe, layer (40 nm thick) the first mask with three parallel slits was turned 90° with respect to its orientation during deposition of the bottom Fe layer.

To fix the magnetization of the top ferromagnetic layer it was either covered by a 25 nm thick film of Cr, or it was made thicker compared to the first ferromagnetic layer. Film thickness during the deposition was controlled by quartz crystal monitor. Deposition speed was 0.6 Å/min for MgO and 3 Å/min for both Fe layers. As a result an array of 3x3 tunnel junctions was formed. Quality of the junctions was evaluated by their resistance. For a 3 nm thick MgO layer the resistance of each of 9 MTJ-structures was varying from few kOhm up to MOhm depending on the parameters of the tunneling barrier.

An important factor for the transport measurements of the fabricated MTJ structures was the proper selection of the magnetic field sweep speed.[10] For this we were considering the MTJ-structure time constant $\tau_{MTJ} = r_{ad}c$ ($r_{ad}$ – dc resistance, $c$ – MTJ capacitance) and using it to limit the time between the magnetic field sweep steps. Magnetic field was applied in the plane of the film structure and was swept in the range of ±500 Oe with the speed of 150 Oe/min. Current wires were attached to the sample by a conductive silver paste. Measurements of the electric and magnetic properties were performed using the reference current source and the digital multimeters with high internal resistance (up to 60 MOhm). To investigate the nature of the conductivity of fabricated junctions their I-V characteristics were studied.

In Fig. 1 we show a typical I-V curve of the Fe(30nm)/MgO(3nm)/Fe(20nm)/Cr(25nm) MTJ-structure with the interlayer resistance of 1.8 kOhm. Measured dependence shows a semiconductor-like conductivity in the range 0-1 V. With the time and also after every measurement cycle the electric properties of the MTJ-structures were degrading. The interlayer resistance of the tunnel junction eventually experienced a significant drop (Fig. 2).

Judging by the exponential behavior of the curve typical to the diffusion-related dependence it is possible to assume that observed degradation is due to diffusion of the conductive impurities. We estimate the average diffusion coefficient through the MgO layer (thickness d = 3 nm) to be $D \approx 3 \cdot 10^{-3}$ nm$^2$/s. This value is more than ten orders of magnitude larger than the volume diffusion coefficient of metals in MgO.[11] As it is shown in Ref. 12 the increased diffusion coefficients are expected for the nano-sized multilayer structures.

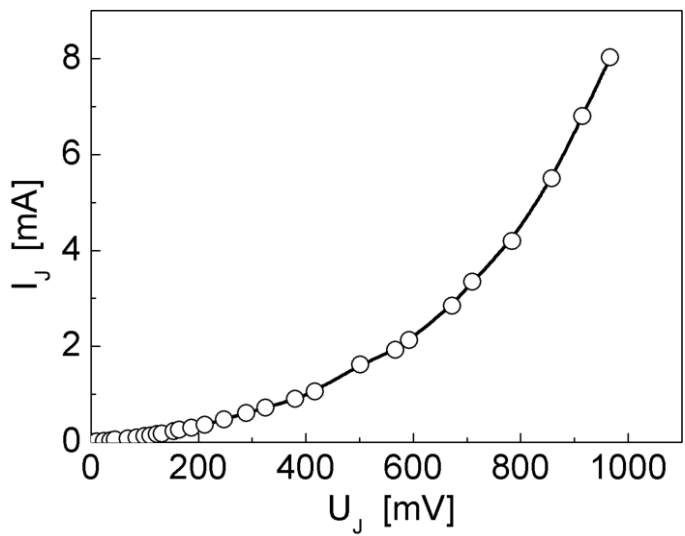

Fig. 1. I-V characteristic of the MTJ-structure with 1.8 kOhm interlayer resistance.

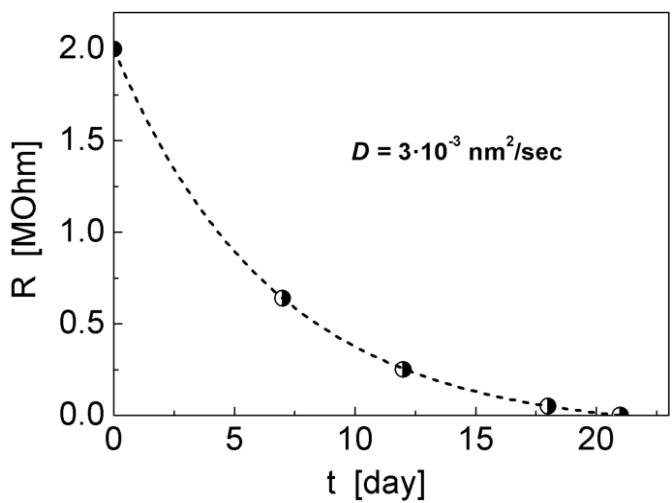

Fig. 2. Degradation of the MTJ properties in time. $D$ is the estimated diffusion coefficient

Besides, the diffusion factor of the MTJ degradation can be aggravated by the interface roughness on both sides of the MgO spacer. Another factor that can contribute to the degradation of the junction is the presence of easily diffusing impurities, in particular carbon and nitrogen, intercalated into the spacer during deposition under the low-vacuum ($10^{-4}$ Torr) conditions. Carbon presence is an inherent factor for the film deposition technology using a diffusion vacuum pump. And one more factor which can possibly lead to the MTJ degradation is the shortcoming of the fabrication technique resulting in the nonuniformity of the barrier layer. The transparency of such barrier is defined by the sparsely distributed regions of small barrier thickness ("pinholes") and experiences giant mesoscopic fluctuations as a function of voltage or magnetic field $H$.[13]

Therefore in order to suppress the MTJ degradation processes a special attention was devoted to smoothing of the interfaces and increase of the MgO layer thickness from 3 nm to 7 nm. The interface smoothing was achieved by thermal stimulation of the surface diffusion of the metal atoms before the deposition of MgO film. As a result a matrix of nine Fe(30nm)/MgO(7nm)/Fe(80nm) tunnel junctions was fabricated having the interlayer resistance of 5-10 MOhm. When measuring such high-resistance junctions in order to avoid their shunting by the multimeter we used an additional resistor ($R_{add}$) connected in series with the MTJ. This resistor was selected to be $R_{add}$ (5 MOhm) << $R_{MTJ}$ for the I-V measurements and $R_{add}$ (2.5 GOhm) >> $R_{MTJ}$ for the TMR measurements. Theoretically the thickness of the MTJ barrier layer can be increased up to the values commensurable with the de Broglie wavelength $\lambda$ of the electron in order to retain the manifestation of their wave properties in the form of tunneling. For the voltage $U$ and temperature $T$ it can be defined from the next equation:[14]

$$\lambda = \frac{2\pi\hbar}{\sqrt{2m_0 E}}, \qquad (1)$$

where $E = kT + eU$. In our case (at $T = 300$ K) $\lambda \sim 10$ nm, which is commensurate with the MgO barrier layer thickness ($d = 7$ nm). The additional resistor provided a small value of the voltage drop across the junction (2 mV) and, correspondingly, current of the order of 1 nA. Using low currents makes it possible to detect finer effects, which are otherwise masked by higher currents. The transparency of the potential barrier can be increased by increasing the temperature, but it will in turn result in the decrease of spin polarization due to an elevated scattering of the electrons. The average transparency coefficient $K$ of such rectangular potential barrier with the height V is:[15]

$$K = K_0 e^{-2a/\hbar \sqrt{2m(V-E)}}, \tag{2}$$

where $a$ – is the width of the barrier equal to the thickness of the insulator (7 nm), $m = 9,1 \cdot 10^{-28}$ g – is the mass of an electron, $E$ – is the energy of an electron, $K_0 \approx 1$. Based on the electric parameters of the MTJ structure the transparency of the barrier is estimated to be 0,1%. From (2) the height of the barrier is $V = 0,02$ eV. From the estimated value we conclude that the barrier layer in our junctions is a doped insulator having semiconductor properties (Fig. 1). It should be noted that for the pure MgO in a similar MTJ structure the height of the potential barrier is on the order of few eV.[16] Therefore, the transparency $K$ can be controlled by changing those barrier layer properties, which are related to the parameter $V$ in (2). For example, the change of the band structure of the insulator in the presence of carbon at the interface with the ferromagnetic layer was observed in Ref. 17.

Figure 3 shows the I-V characteristics of the Fe(30nm)/MgO(7nm)/Fe(80nm) MTJ-structure for the voltage swept in the range 0-3 V. On the initial part (0 to 1 V) of the direct 0 – 3 V branch (Fig. 3a) the curve has nonlinearity typical to semiconductors, similar to the one in Fig. 1. Increase of the voltage may activate the electro-diffusion processes. Impurity atoms, in particular carbon, are moving from one metal-insulator interface towards the other insulator-metal interface. As a result, in the MgO layer with the initially uniformly distributed impurities a concentration gradient starts to form at the interfaces. In this case according to Ref. 18 the presence of impurities in the insulator forms additional energy levels in the band gap, which lead to the appearance of excitons: electron-hole pairs. Thus the re-distribution of impurities and formation of an impurity-rich region at one interface and impurity-poor region at the other results in creation of the energy band structure similar to the one in the tunnel p-n junction. This phenomenon is manifested by the appearance of the region with the negative differential resistance on the reverse branch of the I-V curve (Fig. 3a). Repeated I-V measurement of the same junction (Fig. 3b) shows the negative differential resistance already on both direct (0 to 3 V) and reverse (3 to 0 V) branches.

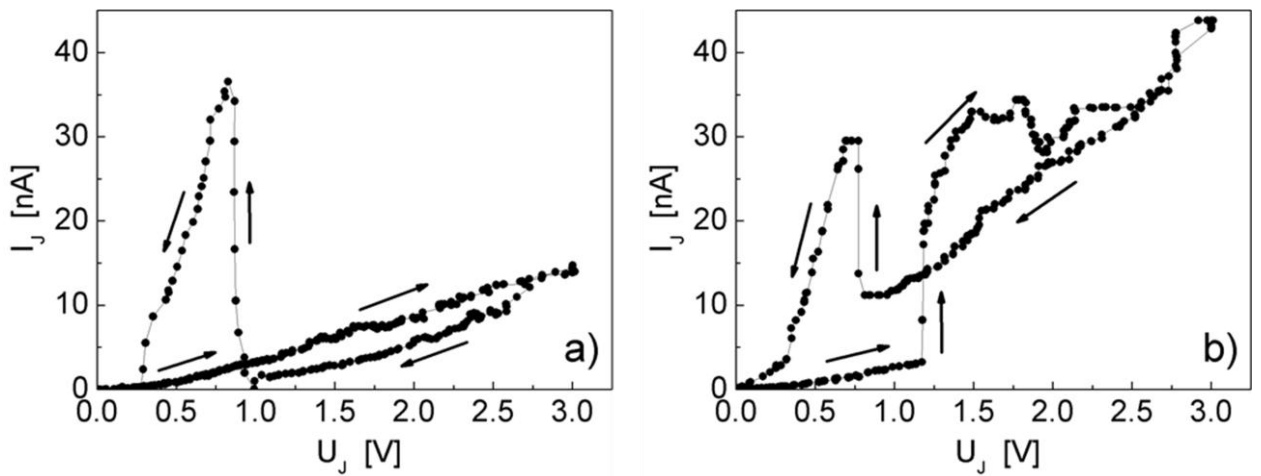

Fig. 3 I-V characteristics of the Fe(30nm)/MgO(7nm)/Fe(80nm) MTJ-structure measured a) initially and b) repeated. Arrows indicate the direction of the voltage sweep.

As was pointed out in the theoretical study (Ref. 8) the formation of the negative differential resistance in the magnetic/insulator/magnetic (M/I/M) structure requires the existence of the valence band in the insulating layer and a certain position of the Fermi level in respect to the band gap of the insulator. In our case such conditions are achieved by introduction of the impurities and their diffusive re-distribution in the insulator.

It was found that properties of the fabricated MTJ-structures are affected by the value of external magnetic field. Results of this study are presented in Fig. 4.

### III. DISCUSSION

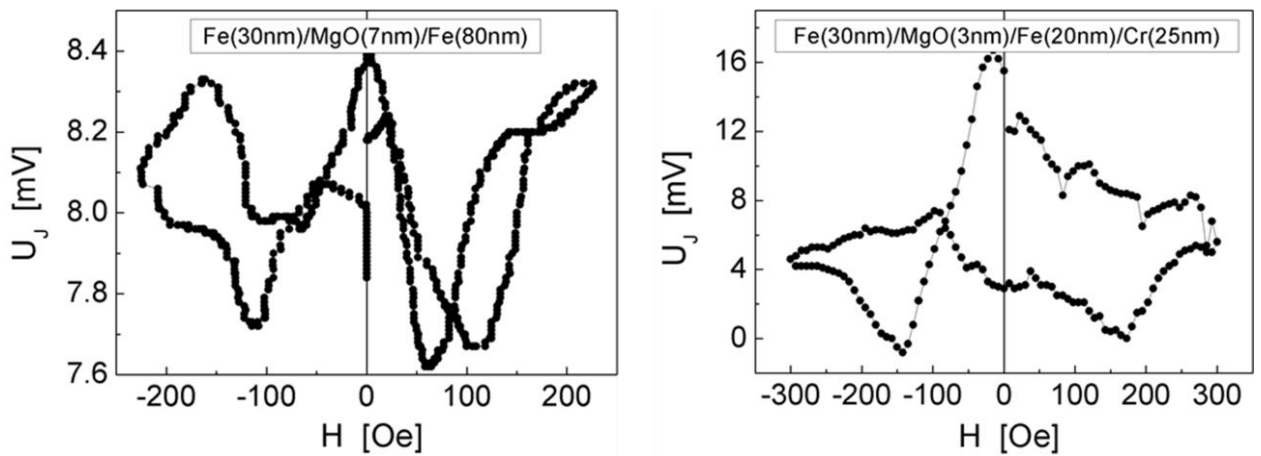

Fig. 4 Magnetic field dependence of the voltage over the MTJ-junction in two different structures (difference in the approach to fix magnetization of one of the ferromagnetic layers and in the thickness of MgO layer)

Magnetic field has an influence on the properties of ferromagnetic films as well as on the electrons themselves. In the thin-film MTJ-structure placed in the longitudinal magnetic field the flowing tunnel current $I_t$ consists of the un-polarized current component $i_n$ and spin-polarized current $i_s$. Spin-polarized current between two ferromagnetic layers increases with the increase of the magnetic field as the magnetization of each layer aligns in the field direction. At the same time with the increase of the magnetic field the Lorentz force acting on the electrons bends their trajectory, the distance electrons travel through the layers increases, which in turn leads to the increase of their scattering. Besides, taking into account the semiconductor type of conductance in these structures, there is a possibility for the additional contribution from the Hall-effect related component of the current $i_h$.

Based on these considerations we now can explain the behavior of the curve presented in Fig. 4a. At the initial moment (H = 0), due to chaotic orientation of the magnetic moments in each ferromagnetic layer caused by their polycrystalline structure, the amount of spin-polarized electrons is statistically negligible. In this case the tunnel current $I_t$ through the MTJ is also minimal and is determined only by the ohmic resistance of the junction.

As the magnetic field increases (H>0) the magnetization vectors of the crystallites in each magnetic layer gradually align along the field and these leads to the increase of the spin current and, as a result, to the decrease of the voltage drop across the MTJ. When the magnetic field reaches saturation field H ≈ $H_S$ for the Fe at about 100-120 Oe,[19] the tunnel current reaches its maximal value. Further increase of the magnetic field leads to the increased scattering of electrons. This results in the decrease of the current through the junction and to the rise of magnetoresistance. As the magnetic field is decreased from the maximum value $H_M$ its influence on the trajectory of the electrons diminishes. In Fig. 4a this case corresponds to the minimal voltage across the MTJ at H ≈ $H_S$. With the further decrease of the magnetic field below $H_S$ its influence on the magnetization of the crystallites in magnetic layers is also decreasing resulting in the increase of the voltage across the MTJ at 0 < H < $H_S$. With the reverse of the magnetic field direction across the H=0 the behavior of the curve is almost symmetric in respect to the case of the positive fields and can be explained in the same way. Some asymmetry of the curve shape may be explained by possible misalignment of the film plane and magnetic field direction. Another possible contribution to the asymmetry may be connected with the change of the total current due to the change of the Hall component with the change of the field polarity. This probably can explain the asymmetry between the minima at saturation fields «+$H_S$» and «-$H_S$».

Another important factor that influences the magnetic properties of the films is their thickness. In particular, as it was shown in Ref. 20, with the increase of the magnetic film thickness its saturation field tends to increase. In structures like Fe/MgO/Fe the increase of one of the magnetic layers thickness is used to "fix" its magnetization. In our case (Fig. 4a) this also results in the difference between the position of the voltage minima around H = 100 Oe during increase and decrease of the magnetic field.

Separate attention should be given to the explanation of the zero and negative values of the voltage in the vicinity of |$H_S$| (Fig. 4b). In this case it is required to take into account the next factors, which are not present in the most studies of the MTJ nanostructures:

1. The current flowing through the MTJ-structure is on the order of 1 nA, in the contrast to commonly used currents on the order of milliamps. This allows us to distinguish much finer effects that are not masked by large currents.

2. The size of the MTJ-cells in our case is about 200x200 μm, which is about $10^3$ times larger than the typical cell size in the most of the MTJ-structure studies.

3. Given such dimensions of the MTJ film structure and its possible slight misalignment relative to the direction of the magnetic field it is necessary to consider the appearance of the planar Hall effect (PHE). The PHE voltage in this case is defined as:[21,22]

$$V_{PH} = \frac{I_x}{d}(\rho_\parallel - \rho_\perp)\sin\theta\cos\theta, \qquad (3)$$

where $V_{PH}$ – PHE voltage in the direction (along Y axis) transverse to the direction (along X axis) of the current $I_x$; $d$ – film thickness; $\theta$ – angle between the magnetization direction and current; $\rho_{\parallel}$ and $\rho\perp$ – resistivity of the film in the direction parallel and perpendicular to the magnetization direction, correspondingly: $(\rho_{\parallel} - \rho\perp) \sim M^2$.

As it follows from Eq. (3) the PHE voltage is related to the magnetization of the film. For the MTJ-structure under discussion the estimated value of the $V_{PH}$ is about $2 \cdot 10^{-9}$ V and cannot significantly affect the measured voltage across the MTJ, which is about two orders of magnitude larger (Fig. 4b).

Apparently, in order to explain the negative values it is necessary to take into account the interplay of all electro-physical phenomena in the junction. The influence of the dimensional and structural nonuniformities of the film may also be significant. It leads to a nonuniform distribution of the potential over the interfaces, similar to what was shown for the quasi-two-dimensional electron gas.[23] Redistribution of the potential over the interface under the influence of magnetic field may result in a significant change of the measured signal. The applied current $I$ give rise to the potential difference in the measurement part of the circuit.[24] However, because of the nonuniformity of the potential, the potential difference between the FM films measured at the edges of the sample may change sign. At the same time the applied current will not change sign. Under the conditions described above it may lead to the reversal of the measured potential difference $U_H$. As shown in Fig. 4b the maximal manifestation of the effect is observed in the fields close to saturation of the ferromagnetic material and is followed by the minimal voltage drop over the MTJ.

Taking into account the fact that the voltage drop over the junction is proportional to its resistance in the changing magnetic field and following the approach in Ref. 25 the value of the TMR for the curve in Fig. 4a can be estimated as:

$$TMR = \frac{U_{ap} - U_p}{U_p} \cdot 100\% = 10.5\% , \qquad (4)$$

while the average spin-polarization coefficient for such structure is determined as

$$P \cong 1/\sqrt{\frac{2}{TMR}+1} \cong 31\% \qquad (5)$$

When the magnetization of one of the Fe films is fixed by Cr (Fig. 4b) the minimum on the direct branch of the MR = f(H), connected with the saturation of Fe, is not observed due to existence of the antiferromagnetic component. With the decrease of the field (after magnetizing the Fe/Cr system up to 300 Oe) the magnetic properties of Fe become dominant with the characteristic minimum close to ~150 Oe. With the further decrease of the field and change of its polarity the linear increase of the voltage drop at the junction is observed. Similar to the previous case the resulting dependence is slightly asymmetric. Around the saturation field $\pm H_S$ the voltage drop across the junction tends to zero. From Eq. (4) and (5) it follows that for this sample at the saturation fields the TMR→∞ and the spin-polarization coefficient P→1.

## IV. CONCLUSIONS

We have shown the possibility of fabrication of the magnetic tunnel junctions by depositing thin film structures on the polycrystalline substrates. The stability of the MTJ properties was achieved by the increase of the dielectric spacer thickness and smoothening of the interface between Fe and MgO layers. Increase of the barrier layer thickness and change of its electrophysical properties leads to a significant decrease of the TMR of the MTJ. The control of the transparency and conductivity of the MTJ potential barrier can be achieved through dosing of the amount and type of introduced impurities. When a potential difference is applied across the MTJ with impurities in the barrier layer there is a possibility of formation of the energy structure as in p-n junction having N-type I-V characteristics similar to the tunneling diode. Appearance of the negative differential resistance may be related to the formation of the excitons in the modified MgO layer. Obtained results will be useful in the development of various types of spintronic devices, and in particular MRAM devices containing MTJ and tunneling diodes.[26]

**ACKNOWLEDGEMENTS**

The work was partly accomplished in the framework of the STCU project N4137.

**References**

[1] M. Bowen, V. Cros, F. Petroff, A. Fert, C. Martinez Boubeta, J. L. Costa-Kramer, J. V. Anguita, A. Cebollada, F. Briones, J. M. de Teresa, L. Morellon, M. R. Ibarra, F. Guell, F. Peiro, A. Cornet, Appl. Phys. Lett. **79**, 1655 (2001).

[2] J. Faure-Vincent, C. Tiusan, E. Jouguelet, F. Canet, M. Sajieddine, C. Bellouard, E. Popova, M. Hehn, F. Montaigne, A. Schuhl, Appl. Phys. Lett. **82**, 4507 (2003).

[3] S. Yuasa, T. Nagahama, A. Fukushima, Y. Suzuki and K. Ando, Nature Materials **3**, 868 (2004).

[4] J. Mathon and A. Umerski, Phys. Rev. B **63**, 220403 (2001).

[5] S. Ahmad, A. Ibrahim, R. Alias, S. M. Shapee, Z. Ambak, S. Z. Zakaria, M. R. Yahya, and A. F. A. Mat, AIP Conf. Proc. **1217**, 442 (2010).

[6] Chin Y. Poon, B. Bhushan, Wear **190**, 89 (1995).

[7] R. Koch, M. Weber, E. Henze, K.H. Rieder, Surf. Sci. **331-333**, 1398 (1995).

[8] T.A. Khachaturova and A.I. Khachaturov, J. Exp. Theor. Phys. **107**, 864 (2008).

[9] A. Vedyaev, D. Bagrets, A. Bagrets, B. Dieny, Phys. Rev. B **63**, 064429 (2001).

[10] R. W. Chantrell, G. N. Coverdale, K. O'Grady, J. Phys. D: Appl. Phys. **21**, 1469 (1988); M. P. Sharrock, IEEE Trans. Magn. **26**, 193 (1990); M. El-Hilo, A.M. de Witte, K. O'Grady, R.W. Chantrell, J. Magn. Magn. Mater. **117**, L307 (1992).

[11] B. J. Wuensch and T. Vasilos, J. Chem. Phys. **42**, 4113 (1965).

[12] E. N. Zubarev, Phys. Usp. **54**, 473 (2011).

[13] M. E. Raikh and I. M. Ruzin, Sov. Phys. JETP **65**, 1273 (1987).

[14] L. de Broglie: Theses de Doctorat (Masson, Paris 1924).

[15] E. O. Kane, in *Tunneling Phenomena in Solids*, edited by E. Burstein and S. Lundqvist, Plenum, New York (1969).

[16] M. E. Eames and J. C. Inkson, Appl. Phys. Lett. **88**, 252511 (2006).

[17] C. Tiusan, F. Greullet, M. Hehn, F. Montaigne, S. Andrieu and A. Schuhl, J. Phys.: Condens. Matter **19**, 165201 (2007).

[18] L.V. Keldysh, Phys. Stat. Sol. **164**, 3 (1997).

[19] M. Prutton, *Thin Ferromagnetic Films* (Butterworths, London, 1964).

[20] Y. K. Kim, M. Oliveria, J. Appl. Phys. **74**, 1233 (1993).

[21] T. R. McGuire, R. I. Potter, IEEE Trans. Magn. **11**, 1018, (1975).

[22] D. G. Stinson, A. C. Palumbo, B. Brandt, M. Berger, J. Appl. Phys. **61**, 3816 (1987).

[23] A. Shik, HAIT J. Sci. Eng. **1**, 470 (2004)

[24] T. Kiyomura, Y. Maruo, and M. Gomia, J. Appl. Phys. **88**, 4768 (2000)

[25] Jian-Gang Zhu and Chando Park, Materials Today **9**, 36 (2006).

[26] T. Uemura, S. Honma, T. Marukame and M. Yamamoto, Jpn. J. Appl. Phys. **43**, L44 (2004).